\documentclass[10pt,letterpaper]{article}
\usepackage[top=0.85in,left=2.75in,footskip=0.75in]{geometry}

\usepackage{amsmath,amssymb}

\usepackage{changepage}

\usepackage{textcomp,marvosym}

\usepackage{cite}

\usepackage{nameref,hyperref}

\usepackage[right]{lineno}
\usepackage{xcolor}

\usepackage[nopatch=eqnum]{microtype}
\DisableLigatures[f]{encoding = *, family = * }

\usepackage[table]{xcolor}

\usepackage{array}

\newcolumntype{+}{!{\vrule width 2pt}}

\newlength\savedwidth

\raggedright
\usepackage[aboveskip=1pt,labelfont=bf,labelsep=period,justification=raggedright,singlelinecheck=off]{caption}

\makeatletter
\renewcommand{\@biblabel}[1]{\quad#1.}
\makeatother

\usepackage{lastpage,fancyhdr,graphicx}
\usepackage{epstopdf}
\fancyheadoffset[L]{2.25in}
\fancyfootoffset[L]{2.25in}

\usepackage[numbers,sort&compress]{natbib}

\usepackage{booktabs}

\usepackage{rotating}

\usepackage{adjustbox}

\begin{document}
\vspace*{0.2in}

\begin{flushleft}
{\Large
\textbf\newline{Tool Choice Matters: Evaluating edgeR vs. DESeq2 for Sensitivity, Robustness, and Cross-Study Performance} 
}
\newline
\\
Mostafa Rezapour, MSc, MA, PhD \textsuperscript{1*}
\\
\bigskip
\textsuperscript{1} Wake Forest Institute for Regenerative Medicine (WFIRM), Wake Forest University School of Medicine, Winston-Salem, NC 27101, USA
\\
\bigskip

* \href{mailto:Mostafa.Rezapour@wfusm.edu}{Mostafa.Rezapour@wfusm.edu}
\end{flushleft}

\section*{Abstract}
Differential gene expression (DGE) analysis is foundational to transcriptomic research, yet tool selection can substantially influence results. This study compares two widely used DGE tools, \texttt{edgeR} and \texttt{DESeq2}, using real and semi-simulated bulk RNA-Seq data sets, mostly from human patients, spanning viral infection, bacterial infection, and fibrotic conditions. We evaluated tool performance across four dimensions: (1) sensitivity to sample size and robustness to outliers; (2) classification performance of uniquely identified gene sets within the discovery dataset; (3) pathway-level concordance of significant DEG sets; and (4) generalizability of tool-specific gene sets across independent studies. First, using Bonferroni-adjusted $p$-value $< 0.05$ and absolute log$_2$ fold change greater than 1 (i.e., $|\log_2 \text{FC}| > 1$) as significance criteria, repeated subsampling showed that \texttt{DESeq2} generally identified more Differentially Expressed Genes (DEGs) than \texttt{edgeR} at smaller sample sizes, while the tools became more concordant as sample size increased. Both tools showed similar responses to simulated outliers, with Jaccard similarity decreasing as more swapped samples were introduced. Second, classification models trained on tool-specific genes showed that \texttt{edgeR} achieved higher F1 scores in 9 of 13 contrasts and more frequently reached perfect or near-perfect precision. Third, Hallmark and KEGG pathway enrichment analyses showed that many contrasts retained substantial pathway-level agreement between tools, although selected contrasts still showed tool-specific enriched pathways. Finally, in cross-study validation using four independent SARS-CoV-2 datasets, \texttt{edgeR}-specific genes yielded higher AUC, precision, and recall in held-out datasets, with some test cases achieving perfect separation. Overall, our findings show that \texttt{DESeq2} may identify more DEGs under stringent thresholds, whereas \texttt{edgeR} often yields more conservative, predictive, and generalizable gene sets. These findings emphasize that DGE tool choice should be guided not only by DEG yield, but also by the downstream reproducibility, predictive value, and biological interpretability of the resulting gene sets.


\section*{Introduction}
Differential gene expression (DGE) analysis is a foundational method in transcriptomics and systems biology, supporting discoveries ranging from disease biomarkers to therapeutic targets \citep{conesa2016survey, wang2009rna, oshlack2010rna, rezapour2025assessing, rezapour2024comparative, rezapour2024analysis, rezapour2024identifying, rezapour2024machineA, rezapour2024machineB, ge2018idep, subramanian2020multi}. The introduction of high-throughput RNA sequencing (RNA-seq) has enabled more accurate and comprehensive gene expression profiling than microarrays \citep{wang2009rna, mortazavi2008mapping, marioni2008rna}. Among RNA-seq methods, bulk RNA-seq remains a widely adopted approach for profiling population-level expression due to its cost-efficiency and broad applicability \citep{conesa2016survey}.

\texttt{edgeR} \citep{robinson2010edger} and \texttt{DESeq2} \citep{love2014moderated} are among the most widely used tools for bulk RNA-seq differential expression analysis. Both rely on the negative binomial distribution to model count data, but they differ in key methodological components: normalization, dispersion estimation, and statistical inference. Specifically, \texttt{edgeR} employs trimmed mean of M-values (TMM) normalization \citep{robinson2010scaling}, tagwise dispersion estimation using empirical Bayes methods \citep{mcCarthy2012differential}, and quasi-likelihood F-tests for statistical testing \citep{lund2012detecting}. In contrast, \texttt{DESeq2} uses median-of-ratios normalization, applies empirical Bayes shrinkage to both dispersion and log$_2$ fold-change estimates, and conducts hypothesis testing using Wald tests \citep{love2014moderated}. \texttt{DESeq} \citep{anders2010differential}, the predecessor to \texttt{DESeq2}, was developed to provide conservative inference by stabilizing dispersion estimates through information sharing across genes.

Several studies have conducted systematic evaluations of RNA-seq DGE tools, including \texttt{edgeR}, \texttt{DESeq}, and \texttt{DESeq2}. Seyednasrollah et al. \citep{seyednasrollah2015comparison} compared eight DGE pipelines using mouse and human RNA-seq datasets with different replicate numbers. They showed that tool choice can substantially affect DEG yield, consistency across subsampled datasets, estimated false discoveries in mock comparisons, and overall agreement among methods. Their work also emphasized that false-discovery behavior depends on sample heterogeneity and that results from very small replicate numbers should be interpreted cautiously. Zhang et al. \citep{zhang2014comparative} compared \texttt{Cuffdiff2}, \texttt{DESeq}, and \texttt{edgeR} across technical and biological replicates, sequencing depth, and balanced versus unbalanced library sizes. They found that biological replication was more influential than sequencing depth for reliable DEG detection and that \texttt{edgeR} recovered more true positives in some settings, but with a higher risk of false positives than \texttt{DESeq}.

More recent studies have extended these concerns to current methods and larger datasets. Stupnikov et al. \citep{stupnikov2021robustness} directly compared \texttt{DESeq2}, \texttt{edgeR}, voom + limma, EBSeq, and NOISeq using breast cancer RNA-seq datasets and fixed count matrices. They evaluated robustness to library-size perturbation, filtering strategy, sample-size reduction, expression level, and downstream gene ontology results. Their results showed that DGE outputs were strongly method-dependent, that robustness patterns became more dataset-dependent when sample size was reduced, and that \texttt{edgeR} was more robust than \texttt{DESeq2} in their framework. Liu et al. \citep{liu2021three} provided a protocol-based comparison of limma, \texttt{edgeR}, and \texttt{DESeq2} using TCGA cholangiocarcinoma data and showed that standard workflows applied to the same dataset and threshold can produce partly overlapping but non-identical DEG sets.

False-positive behavior has also been examined more directly in recent work. Li et al. \citep{li2022exaggerated} evaluated \texttt{DESeq2}, \texttt{edgeR}, limma-voom, NOISeq, dearseq, and the Wilcoxon rank-sum test in large human population-level RNA-seq datasets from GTEx, TCGA, and an immunotherapy study. Using permutation-based negative controls and semi-synthetic datasets, they reported that \texttt{DESeq2} and \texttt{edgeR} could fail to control actual FDR in some large-sample settings and that some spurious DEGs had large estimated fold changes and biologically plausible immune-related enrichment. In a separate evaluation, Li et al. \citep{li2022evaluation} compared eight R/Bioconductor methods across negative-binomial and log-normal simulation settings, equal and unequal library sizes, and sample sizes of 3, 6, and 12 per group. They showed that small sample size produced unstable FDR control and limited method overlap, while performance generally improved as sample size increased; they also found that method ranking depended on the assumed distribution of the count data.

Cui et al. \citep{cui2021super} further addressed type I error control in the development of \texttt{super-delta2}, a method designed for multi-group RNA-seq comparisons. In simulations comparing \texttt{super-delta2} with limma/voom, \texttt{edgeR}, and \texttt{DESeq2}, they reported that \texttt{super-delta2} was the only method that controlled type I error at the nominal level across all simulation settings. Their real-data analysis also showed that \texttt{DESeq2} and \texttt{edgeR} identified more DEGs than \texttt{super-delta2}, a pattern they interpreted as potentially reflecting weaker error control. In addition, they showed that \texttt{edgeR} and \texttt{DESeq2} were more sensitive to filtering choices and that an apparent \texttt{DESeq2}-specific signal could be driven by outliers. Although their work focused on multi-group comparisons, it reinforces the broader concern that DEG yield, model assumptions, outliers, and filtering choices can strongly influence DGE results.

Despite these important contributions, several gaps remain. First, many comparisons rely heavily on simulated or on a small number of benchmark datasets, which may not reflect the diversity of biological signal, tissue source, organism, and study design encountered in applied transcriptomic studies. Second, prior work has focused mainly on statistical operating characteristics such as FDR control, type I error, power, DEG count, or concordance between methods. These metrics are essential, but they do not directly address whether tool-specific genes capture biologically useful signal. Third, most evaluations stop at the DEG list itself and do not test whether genes uniquely identified by one tool can separate biological groups within the discovery dataset, whether they preserve downstream pathway interpretation, or whether they generalize to independent datasets. Fourth, several earlier comparisons evaluated older versions of \texttt{DESeq} or earlier implementations of \texttt{edgeR} and \texttt{DESeq2}, whereas current workflows may behave differently because of updated dispersion estimation, filtering, shrinkage, and testing procedures.

The present study was designed to complement these prior benchmarks by shifting the focus from DEG detection alone to the downstream biological utility and reproducibility of tool-specific gene sets. We compare current implementations of \texttt{edgeR} (v4.4.2) and \texttt{DESeq2} (v1.46.0) using a diverse collection of real-world, biologically annotated bulk RNA-seq datasets spanning viral infection, bacterial infection, and fibrotic lung disease in human and nonhuman primate systems. Our evaluation framework spans four levels: (1) sensitivity to sample size and robustness to controlled outlier perturbation; (2) classification performance of genes uniquely identified by each tool within the discovery dataset; (3) pathway-level concordance of enriched biological processes; and (4) cross-study generalizability of reproducible tool-specific gene sets across independent datasets. By integrating statistical comparison with classification, pathway enrichment, and external validation, this study asks not only which tool identifies more DEGs, but which tool identifies gene sets that are more stable, biologically interpretable, and transferable across studies.

\section*{Materials and methods}

\subsection*{Datasets overview and preprocessing}

This study was designed to systematically compare two widely used RNA-seq differential expression tools, \texttt{edgeR} and \texttt{DESeq2}, using a diverse collection of publicly available bulk RNA-seq datasets. The selected datasets span viral infection, bacterial infection, and chronic fibrotic disease contexts, including SARS-CoV-2, RSVB, EBOV, Mpox, bacterial pneumonia, influenza, and idiopathic pulmonary fibrosis (IPF). These datasets also include both human and nonhuman primate samples, which enabled evaluation across different biological systems and experimental designs. Each dataset included clearly defined control and treated, infected, stimulated, or diseased sample groups.

Datasets were organized according to the main objective of each analysis stage. First, the RSVB dataset (GSE196134) was used for sample-size sensitivity and outlier-perturbation analyses because it provided a balanced design with 45 unstimulated control samples and 45 RSVB-stimulated samples. Second, five datasets were used to assess the classification performance of tool-specific significant genes within the discovery dataset: Mpox (GSE234118), EBOV (GSE115785), bacterial pneumonia and influenza (GSE161731), IPF (GSE231693), and SARS-CoV-2 (PMC8202013). Third, all biological contrasts used for within-dataset DEG comparison were also used for pathway-level concordance analysis. Finally, four independent SARS-CoV-2 datasets, PMC8202013, GSE152418, GSE161731, and GSE171110, were used for cross-study validation of tool-specific gene sets. 

For each dataset, the available gene-count matrix was downloaded from the corresponding public repository and matched with the accompanying metadata. Metadata fields describing infection status, stimulation condition, disease status, or time point were used to assign samples to control and treated groups. Samples with ambiguous or inconsistent metadata annotations were excluded. Analyses were restricted to protein-coding genes, and duplicated or unsupported gene symbols were removed during formatting. After preprocessing, each matrix was organized with genes as rows and samples as columns. No additional normalization was applied before differential expression analysis, because normalization was performed within the tool-specific \texttt{edgeR} and \texttt{DESeq2} workflows. Table~\ref{tab:dataset_summary} summarizes the datasets used in this study, their sample sources, control and treated groups, and their role in the comparative analysis.

\textbf{RSV dataset preprocessing (GSE196134).}
For the RSV sample-size and outlier-perturbation analyses, we used publicly available RNA-seq data from Gene Expression Omnibus (GEO) accession GSE196134 \citep{anderson2024differential}. This dataset was generated from cord blood mononuclear cells (CBMCs) collected from preterm infants, with gestational ages ranging from 30.4 to 34.1 weeks, and term infants, with gestational ages ranging from 37 to 40 weeks. CBMCs from each infant were cultured under unstimulated control and RSVB-stimulated conditions. RSVB stimulation was performed at a multiplicity of infection (MOI) of 1 for 24 hours, resulting in a balanced comparison between unstimulated and RSVB-stimulated samples.

In the original study, CBMCs were stored after stimulation, and total RNA was extracted using the RNeasy Mini Kit. RNA quantity and quality were assessed using the Qubit RNA high-sensitivity assay, and samples with sufficient RNA yield were processed for library preparation. RNA-seq libraries were prepared using the Illumina TruSeq Stranded mRNA Library Prep kit and sequenced on an Illumina NovaSeq 6000 platform with 2 \(\times\) 150 bp paired-end reads. Raw FASTQ files were quality checked using FastQC, reads were aligned to the human transcriptome reference GRCh37v70 using Bowtie2, and gene counts were generated using HTSeq, as described in the original study \citep{anderson2024differential}. For the present analysis, we used 45 unstimulated samples as the control group and 45 RSVB-stimulated samples as the treated group.

\textbf{Mpox dataset preprocessing (GSE234118).}
For the Mpox dataset, we used publicly available whole-blood RNA-seq data from rhesus macaques (\textit{Macaca mulatta}) infected with mpox virus (MPXV; hMPXV/USA/MA001/2022; Lineage B.1, Clade 2b) from GEO accession GSE234118 \citep{aid2023mpox}. In the original study, adult macaques were challenged through intravenous, intradermal, or intrarectal routes with varying doses of MPXV. Whole-blood samples were collected longitudinally before and after infection. Total RNA was extracted from PAXgene blood RNA tubes using the MagMAX RNA Isolation Kit, followed by globin transcript depletion, cDNA synthesis using the Clontech SMART-Seq v4 Ultra Low Input RNA Kit, library preparation using the Nextera XT DNA Library Kit, and sequencing with 100 bp paired-end reads on an Illumina NovaSeq 6000 platform. Reads were aligned to a composite rhesus macaque and MPXV reference genome using STAR, and transcript-level counts were generated using \texttt{htseq-count}, as described in the original study \citep{aid2023mpox}.

For the present analysis, we used the processed count matrix derived from GEO and organized samples according to the time-point labels provided in the metadata. Baseline samples labeled \texttt{wk-1} were used as the uninfected control group. Post-infection samples were mapped as follows: \texttt{wk0d3} was assigned to Mpox DPI 3, \texttt{wk1d7} to Mpox DPI 7, \texttt{wk1d10} to Mpox DPI 10, and \texttt{wk2} to Mpox DPI 14. These time points were selected because they represent the acute phase of MPXV infection, during which viral load and lesion development were reported to peak around days 7--10 in the original study.

\textbf{Ebola virus dataset preprocessing (GSE115785).}
For the EBOV dataset, we used publicly available whole-blood RNA-seq data from rhesus macaques (\textit{Macaca mulatta}) infected with Ebola virus from GEO accession GSE115785 \citep{cross2018comparative}. In the original study, healthy rhesus macaques were challenged intramuscularly with 1000 PFU of the EBOV Makona C05 isolate. Whole-blood samples were collected before infection and at defined post-infection time points, including days 5 and 7 post infection, as well as at necropsy from terminal animals. RNA was extracted from whole blood, and RNA quality was assessed using the Agilent Bioanalyzer 2100. Ribosomal RNA was depleted using the Illumina TruSeq Stranded RNA Ribo-Zero Gold kit, and sequencing libraries were generated from 500 ng of input RNA. Libraries were sequenced on an Illumina NextSeq 500 platform using paired-end 50 bp reads, with adapter trimming and demultiplexing performed using the Illumina BaseSpace platform, as described in the original study \citep{cross2018comparative}.

For the present analysis, we used the processed gene-count matrix derived from GEO and organized samples according to the time-point labels provided in the metadata. Samples collected before infection were used as the baseline control group and labeled EBOV DPI 0. Post-infection samples were grouped as EBOV DPI 5, EBOV DPI 7, and EBOV NEC, where EBOV NEC denotes samples collected at necropsy from terminal animals. These groups were selected to represent baseline, acute infection, later infection, and terminal disease states.

\textbf{Bacterial pneumonia and influenza dataset preprocessing (GSE161731).}
For the bacterial pneumonia and influenza comparisons, we used publicly available peripheral whole-blood RNA-seq data from GEO accession GSE161731 \citep{mcclain2021dysregulated}. This dataset was generated to characterize peripheral blood transcriptional responses across subjects with acute respiratory infections, including influenza, bacterial pneumonia, and healthy controls. In the original study, total RNA was extracted from whole blood using the PAXgene Blood miRNA Kit, libraries were prepared using the NuGEN Universal mRNA-seq kit with AnyDeplete Globin, and sequencing was performed on an Illumina NovaSeq 6000 platform \citep{mcclain2021dysregulated}. For the present analysis, we used the processed count matrix from GEO and focused on three groups: bacterial pneumonia, influenza, and healthy controls. Healthy controls were used as the baseline group, while bacterial pneumonia and influenza samples were analyzed as separate treated groups. 

\textbf{Fibrotic interstitial lung disease dataset preprocessing (GSE231693).}
For the IPF comparison, we used publicly available bulk RNA-seq data from GEO accession GSE231693 \citep{jia2023interleukin}. This dataset was generated from explanted lung tissue samples collected from patients with idiopathic pulmonary fibrosis, systemic sclerosis-associated interstitial lung disease, and healthy controls. In the original study, total RNA was isolated from lung tissue samples and used to prepare sequencing libraries with the Illumina TruSeq RNA Sample Prep Kit v2. Libraries were sequenced on an Illumina HiSeq 4000 platform, as described by Jia et al. \citep{jia2023interleukin}. For the present study, we focused on the IPF and healthy control groups. IPF samples were analyzed as the diseased group, and healthy lung samples were used as the baseline group. This comparison was included to evaluate whether tool-specific DEG behavior observed in infectious disease datasets also extended to a chronic fibrotic lung disease context.

\textbf{SARS-CoV-2 dataset preprocessing.}
For cross-study validation, we used four independent SARS-CoV-2 bulk RNA-seq datasets: GSE152418, PMC8202013, GSE161731, and GSE171110. These datasets were selected because each included clearly defined SARS-CoV-2-positive and healthy control groups, which enabled consistent differential expression analysis and leave-one-dataset-out validation across independent cohorts.

For GSE152418, we used PBMC RNA-seq data from Arunachalam et al. \citep{arunachalam2020systems}. RNA-seq libraries were generated from cryopreserved PBMCs and sequenced on an Illumina NovaSeq 6000 platform, as described in the original study. For our analysis, the processed count matrix was formatted with gene symbols as row identifiers and samples as columns, and sample labels were standardized as Control and SARS-CoV-2 groups.

For PMC8202013, we used whole-blood RNA-seq data from Bibert et al. \citep{bibert2021transcriptomic}. COVID-19 patients represented a range of respiratory failure severities, including patients not requiring oxygen support, patients requiring oxygen without mechanical ventilation, and intubated patients. In the present study, all SARS-CoV-2-positive samples were analyzed together as the treated group, and healthy individuals were used as the control group. RNA sequencing was performed on whole-blood samples using the Illumina HiSeq 4000 platform, as described in the original study. This dataset was also used in the within-dataset classification analysis.

For GSE161731, we used whole-blood RNA-seq data from McClain et al. \citep{mcclain2020dysregulated}. Total RNA was extracted from whole blood using the PAXgene Blood miRNA Kit, libraries were prepared using the NuGEN Universal mRNA-seq kit with AnyDeplete Globin, and sequencing was performed on an Illumina NovaSeq 6000 platform. The processed count matrix was formatted with gene symbols as row identifiers, and sample labels were standardized as Control and SARS-CoV-2 groups.

For GSE171110, we used whole-blood RNA-seq data from Lévy et al. \citep{levy2021cd177}. Blood samples for immunological and transcriptomic profiling were collected near hospital admission, with most patients requiring intensive care. Total RNA was purified from whole blood using the Tempus Spin RNA Isolation Kit, RNA quantity was measured using the Quant-iT RiboGreen RNA Assay Kit, and RNA quality was evaluated using a Bioanalyzer. Globin mRNA was depleted using the GLOBINclear Kit, libraries were prepared using the TruSeq Stranded mRNA Kit, and sequencing was performed on an Illumina HiSeq 2500 V4 system. For the present analysis, sample labels were standardized as Control and SARS-CoV-2 groups.

To ensure consistency across the manuscript, dataset names are used without GEO or source accession IDs when the context clearly identifies the dataset being discussed. However, in figures and places where multiple related contrasts are compared, dataset contrasts are reported using the format ``accession ID | biological contrast'' to improve clarity and avoid ambiguity. For longitudinal infection datasets, contrast labels also include the day post infection (DPI) or necropsy (NEC) time point when applicable. For example, Mpox day 7 is labeled as GSE234118 | Mpox DPI 7.

\begin{table}[!ht]
\begin{adjustwidth}{-2.25in}{0in}
\centering
\caption{\textbf{Summary of RNA-seq datasets used in this study.} Each dataset was selected to enable systematic comparison of differential gene expression results from \texttt{edgeR} and \texttt{DESeq2} under well-defined baseline and treated, infected, stimulated, or diseased groups.}
\label{tab:dataset_summary}
\renewcommand{\arraystretch}{1.25}
\setlength{\tabcolsep}{3pt}
{\footnotesize
\begin{tabular}{p{1.15in} p{1.35in} p{1.15in} p{1.35in} p{2.05in}}
\hline
\textbf{Dataset} & \textbf{Sample source} & \textbf{Control group} & \textbf{Treated / disease group} & \textbf{Application in current study} \\
\hline

RSVB (GSE196134) \citep{anderson2024differential} &
Cord blood mononuclear cells from preterm and term infants &
Unstimulated CBMCs (n = 45) &
RSVB-stimulated CBMCs at MOI = 1 for 24 hours (n = 45) &
Used for sample-size sensitivity analysis and outlier-perturbation analysis because it provided a balanced control and stimulated design. \\

\hline
Mpox (GSE234118) \citep{aid2023mpox} &
Peripheral whole blood from rhesus macaques infected with MPXV &
Baseline pre-infection samples labeled \texttt{wk-1} &
Post-infection samples at DPI 3, DPI 7, DPI 10, and DPI 14 &
Used for within-dataset DEG comparison, tool-specific gene classification, and pathway-level concordance analysis across acute MPXV infection time points. \\

\hline
EBOV (GSE115785) \citep{cross2018comparative} &
Whole blood from rhesus macaques challenged with EBOV Makona C05 &
Pre-infection baseline samples labeled EBOV DPI 0 &
EBOV DPI 5, EBOV DPI 7, and EBOV NEC samples &
Used for within-dataset DEG comparison, tool-specific gene classification, and pathway-level concordance analysis across acute and terminal EBOV disease states. \\

\hline
Bacterial pneumonia and influenza (GSE161731) \citep{mcclain2021dysregulated} &
Peripheral whole blood from patients with acute respiratory infection and healthy controls &
Healthy controls (n = 16) &
Bacterial pneumonia (n = 23) and influenza (n = 17) &
Used for within-dataset DEG comparison, tool-specific gene classification, and pathway-level concordance analysis in bacterial and viral respiratory infection contrasts. \\

\hline
Fibrotic interstitial lung disease (GSE231693) \citep{jia2023interleukin} &
Explanted lung tissue from patients with IPF, systemic sclerosis-associated interstitial lung disease, and healthy controls &
Healthy lung tissue (n = 18) &
IPF lung tissue (n = 20) &
Used for within-dataset DEG comparison, tool-specific gene classification, and pathway-level concordance analysis in chronic fibrotic lung disease. \\

\hline
SARS-CoV-2 (PMC8202013) \citep{bibert2021transcriptomic} &
Whole blood from COVID-19 patients and healthy controls &
Healthy controls (n = 27) &
COVID-19 patients across respiratory failure severities (n = 103) &
Used for within-dataset classification, pathway-level concordance analysis, and cross-study SARS-CoV-2 validation. \\

\hline
SARS-CoV-2 (GSE152418) \citep{arunachalam2020systems} &
PBMCs from SARS-CoV-2-positive patients and healthy controls &
Healthy controls (n = 17) &
COVID-19 patients with moderate, severe, or ICU-level disease (n = 16) &
Used for cross-study SARS-CoV-2 validation. \\

\hline
SARS-CoV-2 (GSE161731) \citep{mcclain2020dysregulated} &
Whole blood from SARS-CoV-2-positive patients and healthy controls &
Healthy controls (n = 16) &
COVID-19 patients (n = 12) &
Used for cross-study SARS-CoV-2 validation. \\

\hline
SARS-CoV-2 (GSE171110) \citep{levy2021cd177} &
Whole blood from severe COVID-19 patients and healthy donors &
Healthy donors (n = 10) &
Severe COVID-19 patients, mostly ICU cases (n = 44) &
Used for cross-study SARS-CoV-2 validation. \\

\hline
\end{tabular}
}
\end{adjustwidth}
\end{table}

\subsection*{Tool configuration and execution strategy}

To simulate real-world application and usability, we ran both \texttt{edgeR} and \texttt{DESeq2} using their recommended and widely adopted default pipelines. These reflect the typical usage patterns of practitioners who perform standard RNA-Seq differential expression analysis using default guidance from respective documentation and vignettes. The following summarizes how each major methodological component was implemented in \texttt{edgeR} (v4.4.2) and \texttt{DESeq2} (v1.46.0):

\begin{itemize}
  \item \textbf{Model Framework:} 
  In \texttt{edgeR}, a negative binomial GLM is fitted using quasi-likelihood methods. Differential expression (DE) analysis is performed using \texttt{glmQLFit()} followed by \texttt{glmQLFTest()}. In \texttt{DESeq2}, a negative binomial GLM is fitted using \texttt{DESeq()}, and hypothesis testing is conducted using the Wald test.

  \item \textbf{Normalization:} 
  \texttt{edgeR} performs normalization using the \texttt{calcNormFactors()} function, which implements the Trimmed Mean of M-values (TMM) method. \texttt{DESeq2} uses the median-of-ratios method implemented in \texttt{estimateSizeFactors()}.

  \item \textbf{Dispersion Estimation:} 
  \texttt{edgeR} estimates common, trended, and tagwise dispersions using \texttt{estimateDisp()}. \texttt{DESeq2} uses \texttt{estimateDispersions()} to compute gene-wise and fitted dispersions.
  
\item \textbf{Statistical Test and Output:} 
\texttt{edgeR} applies the GLM quasi-likelihood F-test via \texttt{glmQLFTest()}. 
In \texttt{DESeq2}, the \texttt{DESeq()} function fits the negative binomial GLM and, under the default setting, performs Wald tests for the model coefficients; likelihood ratio tests are also available as an alternative option. The \texttt{results()} function was then used to extract the requested coefficient estimates, Wald test $p$-values, adjusted $p$-values, and log$_2$ fold changes.

\end{itemize}

\subsection*{Sensitivity to sample size and robustness to outliers}

The first phase of our comparative evaluation aimed to assess how \texttt{edgeR} and \texttt{DESeq2} respond to sample size variation and the presence of outliers, both common challenges in real-world RNA-Seq experiments. To perform this, we selected the RSVB dataset (GSE196134), which contains 90 total samples: 45 control (unstimulated) and 45 RSVB-infected (stimulated). This balanced and large dataset enabled controlled subsampling and simulation of outlier scenarios.

We first applied both tools to the full dataset, contrasting the RSVB-infected group against the control group. Subsequently, we generated three subsampled datasets, each with 20, 10, and 5 samples per group, respectively. Each subsampled dataset was randomly drawn from the original 45 samples per group. This design allowed us to evaluate the sensitivity of each method to reductions in sample size. For each dataset version (full and subsampled), we applied the same analysis pipeline using \texttt{edgeR} and \texttt{DESeq2}. DEGs were defined as those with a Bonferroni-adjusted $p-value <$ 0.05 \citep{armstrong2014use} and an absolute log$_2$ fold change $| \log_2 \text{FC} | > 1$. This stringent threshold was chosen because applying an FDR cutoff resulted in an excessively large number of differentially expressed genes for both tools, which hindered meaningful comparison. Using Bonferroni-adjusted p-values allowed for a more conservative and balanced comparison, which better suited to our goal of evaluating differences between the two methods. To quantify consistency and directional agreement between tools, we introduced the following metric:

\textbf{Definition: }
Let \( A \) and \( B \) be two sets of differentially expressed genes. The ordered pair \( (A, B) \) is interpreted as comparing the set \( A \) against the reference set \( B \). The \emph{Directional Overlap}, or \( \mathrm{DO}(A, B) \), is defined as the proportion of elements in \( B \) that are also found in \( A \), given by
\begin{equation}
\mathrm{DO}(A, B) = \frac{|A \cap B|}{|B|},
\label{eq:directional_overlap}
\end{equation}
which quantifies the extent to which the reference set \( B \) is recovered by the comparison set \( A \). Notably, \( \mathrm{DO}(A, B) \neq \mathrm{DO}(B, A) \) in general due to its asymmetry. For the sample-size analysis, we computed directional overlap between the DEG sets identified by \texttt{edgeR} and \texttt{DESeq2} within each subsampling replicate. For $n=5$, $n=10$, and $n=20$ samples per group, directional overlap values were summarized as mean $\pm$ standard deviation across 20 independent subsampling replicates. For the full dataset ($n=45$ samples per group), directional overlap was computed once using the complete dataset.

We also evaluated the tools' robustness to outliers by introducing controlled sample swaps between the treatment and control groups. Specifically, we generated synthetic outliers by swapping 1 to 5 samples between groups. For instance, in the 1-swap scenario, one control sample was swapped with one RSVB-infected sample, thus injecting one outlier into each group. This process was repeated for 2 through 5 swaps. To ensure statistical robustness and mitigate the effect of random variation in swap choices, each swap level (1 to 5) was independently repeated 20 times. Given the original dataset’s size (45 per group), the proportion of swaps remained appropriate. 

To assess changes introduced by these simulated outliers, we computed the Jaccard Index \citep{fletcher2018comparing} between the original DEG set (from the full, unperturbed dataset) and the DEG set after introducing outliers:

\begin{equation}
J(A, B) = \frac{|A \cap B|}{|A \cup B|},
\label{eq:jaccard_index}
\end{equation}
where, \( A \) represents the set of significant genes identified after injecting outliers, and \( B \) is the original DEG set without outliers. Jaccard similarity provides a symmetric measure of agreement and reflects how much the gene sets overlap before and after contamination with outliers.

\subsection*{Classification performance of uniquely identified gene sets within the discovery dataset}

In the second phase of our analysis, we systematically compared \texttt{edgeR} and \texttt{DESeq2} across multiple datasets to evaluate the concordance and discrepancy in their differential expression results. To evaluate the biological signal of tool-specific genes, we applied this analysis to five representative datasets covering diverse biological conditions: Mpox (\text{GSE234118}), EBOV (\text{GSE115785}), Bacterial and Influenza infection (\text{GSE161731}), Idiopathic Pulmonary Fibrosis (IPF, \text{GSE231693}), and SARS-CoV-2 (\text{PMC8202013}).

For each dataset, we quantified the number of significant genes identified by each tool using a stringent threshold: Bonferroni-adjusted $p-value < 0.05$ and $| \log_2 \text{FC} | > 1$. We then evaluated concordance and divergence between \texttt{edgeR} and \texttt{DESeq2} by reporting the number of genes uniquely identified by each tool, the number of shared significant genes, and the direction-specific Jaccard indices (Equation~\ref{eq:jaccard_index}) for upregulated and downregulated gene sets, where \( A \) and \( B \) represent the respective sets from \texttt{edgeR} and \texttt{DESeq2}. To further examine agreement between tools, we calculated Pearson \citep{cohen2009pearson} and Spearman \citep{de2016comparing} correlation coefficients for the log$_2$ fold changes and Bonferroni-adjusted $p$-values among the common significant genes.

Finally, to assess how effective each tool is in identifying biologically meaningful significant genes, we focused specifically on the genes uniquely identified by each method. In the absence of an external ground truth for differentially expressed genes, one way to evaluate the ``trueness" of these genes is to assess their ability to differentiate between biological groups, in this case, control and treated samples.

For each dataset, we extracted the genes uniquely identified as significant by each tool. Raw count values for these genes were log-transformed using $\log_2(\text{count} + 1)$ without any normalization to preserve the original scale and avoid introducing method-dependent biases. Principal Component Analysis (PCA) \citep{abdi2010principal} was then applied to reduce dimensionality while retaining key variance components. We retained only the first two principal components (PC1 and PC2), which capture the dominant structure in the expression space of the selected genes and help mitigate overfitting while maintaining interpretability.

PC1 and PC2 were used as predictors in a logistic regression classifier \citep{ng2001discriminative} trained to distinguish control from treated samples. Standard classification metrics, accuracy, precision, recall, and F1 score \citep{yacouby2020probabilistic}, were computed to evaluate how well the tool-specific significant genes separated the biological conditions. A higher classification performance suggests that the genes uniquely identified by the corresponding tool are more likely to reflect meaningful biological signal.

To further benchmark and summarize tool performance across all datasets and conditions, we adopted the Dolan-More performance profiling method \citep{dolan2002benchmarking}. This technique evaluates each method's relative performance consistency across datasets using the F1 score as the basis for comparison.

Let $s_{a,i}$ denote the F1 score of method $a$ on dataset $i$, and let $s_i^* = \max_a s_{a,i}$ be the highest F1 score obtained on dataset $i$ across methods. The performance ratio for each method is defined as:

\[
r_{a,i} = \frac{s_i^*}{s_{a,i}} \quad \text{for all } i,
\]
with the convention that $r_{a,i} = \infty$ if $s_{a,i} = 0$. A smaller $r_{a,i}$ indicates better performance, with $r_{a,i} = 1$ signifying that the method achieved the best score on that dataset. The Dolan-More profile for method $a$ is then the cumulative distribution function:

\[
\rho_a(\tau) = \frac{1}{n} \left| \left\{ i \in \{1,\dots,n\} \mid r_{a,i} \leq \tau \right\} \right|,
\]
where $n$ is the total number of datasets. This function quantifies the fraction of datasets where a given method’s performance is within a factor $\tau$ of the best-performing method. A Dolan-More curve that rises quickly and reaches $\rho_a(\tau) = 1$ earlier indicates a more consistently high-performing method. In our context, this analysis provides a global, dataset-agnostic perspective that complements pairwise comparisons and highlights each tool's overall reliability and robustness in identifying biologically meaningful gene sets.

\subsection*{Pathway enrichment concordance analysis}

To determine whether gene-level differences between \texttt{edgeR} and \texttt{DESeq2} translated into differences in downstream biological interpretation, we performed a pathway enrichment concordance analysis using the significant DEG sets identified by each tool. This analysis was added because differences in the number or identity of significant DEGs may not necessarily lead to divergent pathway-level conclusions.

For each dataset contrast and each tool, we used the significant DEG set defined by the same threshold applied in the corresponding DEG analysis. We then performed pathway over-representation analysis using two curated pathway collections: the MSigDB Hallmark gene sets, which summarize coherent biological processes with reduced redundancy \citep{subramanian2005gene, liberzon2015molecular}, and KEGG pathways, which provide manually curated pathway maps for biological systems and molecular interactions \citep{kanehisa2000kegg, kanehisa2021kegg}.

For each pathway collection, enrichment was tested using the hypergeometric test. The background was defined as the union of genes represented in the corresponding pathway collection. Pathway-level \(p\)-values were adjusted using the Benjamini-Hochberg procedure, and pathways with FDR \(< 0.05\) were considered significantly enriched. For each dataset contrast, enriched pathways were classified as \texttt{edgeR}-only, shared by both tools, or \texttt{DESeq2}-only.

To quantify pathway-level agreement between tools, we computed the Jaccard similarity between the enriched pathway sets from \texttt{edgeR} and \texttt{DESeq2}:
\[
J_{\mathrm{pathway}}(E,D)=\frac{|P_E \cap P_D|}{|P_E \cup P_D|},
\]
where \(P_E\) and \(P_D\) denote the sets of significantly enriched pathways derived from \texttt{edgeR} and \texttt{DESeq2} DEG sets, respectively. Higher values indicate greater pathway-level concordance between tools.

\subsection*{Generalizability of tool-specific gene sets across independent studies}

As the final stage of our analysis, we evaluated the generalizability of the significant genes uniquely identified by each tool. While the previous classification-based analysis assessed how well tool-specific genes separated samples within the same dataset in which they were discovered, this section extends the evaluation to a more rigorous cross-dataset framework. Specifically, we asked: to what extent are the unique genes identified by each method in some datasets transferable and predictive in unseen datasets? This helps assess whether the tool-specific genes capture biologically consistent signals that generalize across studies.

We focused this evaluation on four independent bulk RNA-Seq datasets of SARS-CoV-2 infection: GSE152418, GSE161731, GSE171110, and PMC8202013. To ensure consistent gene coverage across datasets, we first aligned all datasets to a common gene set by restricting our analysis to genes shared across all four datasets. For each tool (\texttt{edgeR} and \texttt{DESeq2}), we identified significant genes in each dataset using an FDR threshold of 0.05 and an absolute log$_2$ fold change $| \log_2 \text{FC} | > 1$. Unlike the previous section, where we applied the more conservative Bonferroni correction due to reliance on a single dataset, we used the FDR approach here because this analysis required reproducibility across multiple independent datasets. This cross-study design helps reduce the influence of dataset-specific findings and provides a more appropriate framework for evaluating transferable gene signatures.

We implemented a leave-one-dataset-out cross-study validation strategy using the four SARS-CoV-2 datasets. In each fold, three datasets were used as training datasets and the remaining dataset was used as the held-out test dataset. This procedure was repeated four times, so that each dataset served once as the held-out test dataset. The four folds were: (1) GSE152418 held out, with GSE161731, GSE171110, and PMC8202013 used for training; (2) GSE161731 held out, with GSE152418, GSE171110, and PMC8202013 used for training; (3) GSE171110 held out, with GSE152418, GSE161731, and PMC8202013 used for training; and (4) PMC8202013 held out, with GSE152418, GSE161731, and GSE171110 used for training.

Within each fold, significant genes were first identified separately for each tool in each of the three training datasets. Let \(E_{T_1}\), \(E_{T_2}\), and \(E_{T_3}\) denote the significant genes identified by \texttt{edgeR} in the three training datasets, and let \(D_{T_1}\), \(D_{T_2}\), and \(D_{T_3}\) denote the corresponding significant genes identified by \texttt{DESeq2}. We first defined the reproducible training gene set for each tool as the intersection of significant genes across the three training datasets:
\[
E_{\mathrm{train}} = E_{T_1} \cap E_{T_2} \cap E_{T_3}
\]
\[
D_{\mathrm{train}} = D_{T_1} \cap D_{T_2} \cap D_{T_3}.
\]

This intersection represented the genes reproducibly identified by each tool across independent studies. After defining these reproducible training gene sets, we removed genes shared between \texttt{edgeR} and \texttt{DESeq2}, because the purpose of this analysis was to evaluate tool-specific signals rather than genes identified by both methods. Thus, the \texttt{edgeR}-specific and \texttt{DESeq2}-specific training signatures were defined as:
\[
E_{\mathrm{unique}} = E_{\mathrm{train}} \setminus D_{\mathrm{train}}
\]
\[
D_{\mathrm{unique}} = D_{\mathrm{train}} \setminus E_{\mathrm{train}}.
\]

These tool-specific unique gene sets were then evaluated in the held-out dataset that was not used during gene selection. This design separates gene discovery from testing and evaluates whether genes uniquely and reproducibly identified by each tool in three independent studies can classify samples in an unseen fourth study.

To evaluate the predictive power of these tool-specific unique gene sets, we extracted raw expression values (log-transformed using $\log_2(\text{count} + 1)$, without normalization) for the unique genes from the test dataset. As before, we performed PCA and retained the first two principal components (PC1 and PC2), which capture the dominant variation within each gene set. These components were used to classify control versus SARS-CoV-2–positive samples via a logistic regression model. Classification performance was quantified using metrics including area under the ROC curve (AUC) \citep{hoo2017roc}, precision, and recall.

Note that this approach differs from the within-dataset classification described in the previous subsection. In that approach, genes were selected and evaluated on the same dataset, potentially capturing dataset-specific variance or artifacts. In contrast, the current design separates gene discovery (training) and evaluation (testing) across distinct datasets, which provides a more stringent test of biological generalizability and reproducibility. This mirrors standard machine learning principles by assessing how well tool-specific discoveries made during training generalize to unseen, external data.

\section*{Results}

\subsection*{Sensitivity to sample size and robustness to outliers}

\begin{figure}[!ht]
\centering
\includegraphics[width=0.93\textwidth]{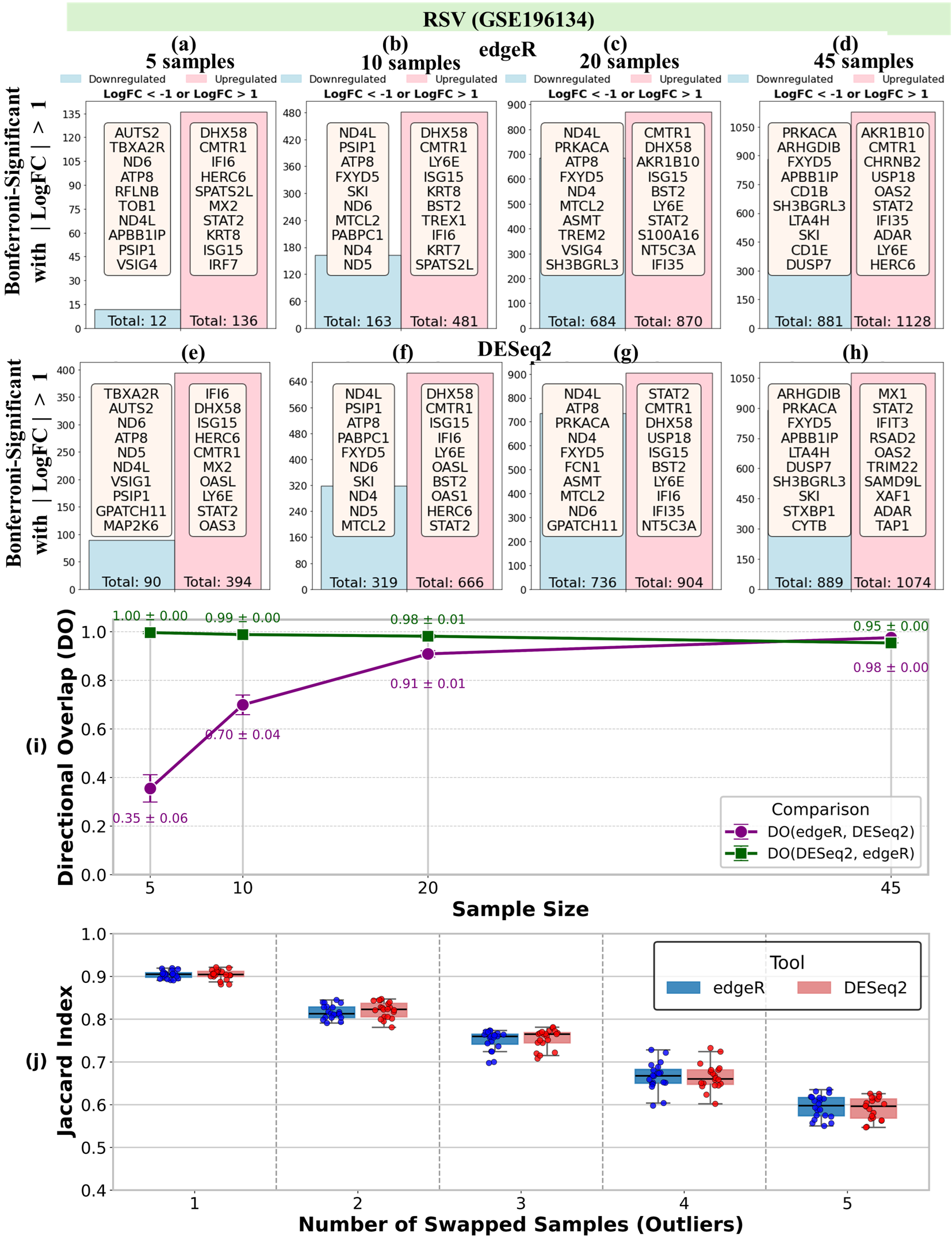}
\caption{\textbf{Sensitivity of edgeR and DESeq2 to sample size and outliers.}
Panels (a--d) show DEG counts and top significant genes for \texttt{edgeR} across 5, 10, 20, and 45 samples per group. 
Panels (e--h) show the same for \texttt{DESeq2}. 
For reduced sample sizes ($n=5$, $n=10$, and $n=20$ per group), panels (a--c) and (e--g) show representative subsampled datasets, while panel (i) summarizes directional overlap across 20 independent random subsampling replicates at each reduced sample size. 
Panel (i) shows mean $\pm$ standard deviation directional overlap between DEG sets identified by the two tools; the full dataset ($n=45$ per group) was analyzed once. 
Panel (j) shows DEG set stability under sample swapping using Jaccard similarity, with 20 independent replicates for each swap level from 1 to 5.}
\label{fig:Main-1}
\end{figure}

Figure~\ref{fig:Main-1} panels (a--h) summarize the DEG discovery performance of \texttt{edgeR} and \texttt{DESeq2} as sample size increases from 5 to 45 samples per group using the RSVB dataset. Panels (a--d) show results for \texttt{edgeR}, and panels (e--h) show results for \texttt{DESeq2}. For the reduced sample-size settings ($n=5$, $n=10$, and $n=20$ per group), the displayed bar plots show representative subsampled datasets, while the directional overlap analysis in panel (i) summarizes 20 independent random subsampling replicates. In the representative subsamples shown, \texttt{DESeq2} identified more DEGs than \texttt{edgeR} at $n=5$ (484 versus 148), $n=10$ (985 versus 644), and $n=20$ (1,640 versus 1,554). In the full dataset ($n=45$ per group), \texttt{edgeR} identified 2,009 DEGs and \texttt{DESeq2} identified 1,963 DEGs. Thus, the difference between tools was most pronounced at smaller sample sizes and decreased as sample size increased.

Panel (i) quantifies directional overlap (DO) between DEGs identified by the two tools at each sample size. For $n=5$, $n=10$, and $n=20$ per group, DO values were computed separately for each of the 20 independent subsampling replicates and are reported as mean $\pm$ standard deviation. At $n=5$, $\mathrm{DO}(\texttt{edgeR}, \texttt{DESeq2}) = 0.35 \pm 0.06$, indicating that edgeR recovered a smaller fraction of DESeq2's DEG set at the smallest sample size, while $\mathrm{DO}(\texttt{DESeq2}, \texttt{edgeR}) = 1.00 \pm 0.00$, indicating that DESeq2 recovered all edgeR DEGs across replicates. At $n=10$, the corresponding values were $0.70 \pm 0.04$ and $0.99 \pm 0.00$, and at $n=20$, they were $0.91 \pm 0.01$ and $0.98 \pm 0.01$. In the full dataset ($n=45$ per group), directional overlap was nearly symmetric, with $\mathrm{DO}(\texttt{edgeR}, \texttt{DESeq2}) = 0.98$ and $\mathrm{DO}(\texttt{DESeq2}, \texttt{edgeR}) = 0.95$. These replicated subsampling results confirm that agreement between tools is lowest at small sample size and increases as sample size grows.

Panel (j) evaluates the tools' robustness to sample contamination by computing the Jaccard similarity between DEG sets derived from the original full dataset and those obtained after introducing 1 to 5 randomly swapped sample pairs between the control and RSVB groups. Each swap level was repeated 20 independent times. Both tools showed a monotonic decline in Jaccard similarity as the number of swapped samples increased, indicating reduced DEG-set stability with increasing contamination. Across swap levels, \texttt{edgeR} and \texttt{DESeq2} showed broadly similar behavior, with \texttt{DESeq2} showing slightly higher median Jaccard similarity at several swap levels. These results indicate that both tools are sensitive to increasing sample-label perturbation, while the magnitude of the difference between tools under this perturbation is modest.

\subsection*{Classification performance of uniquely identified gene sets within the discovery dataset}

\begin{figure}[!ht]
\centering
\includegraphics[width=0.95\textwidth]{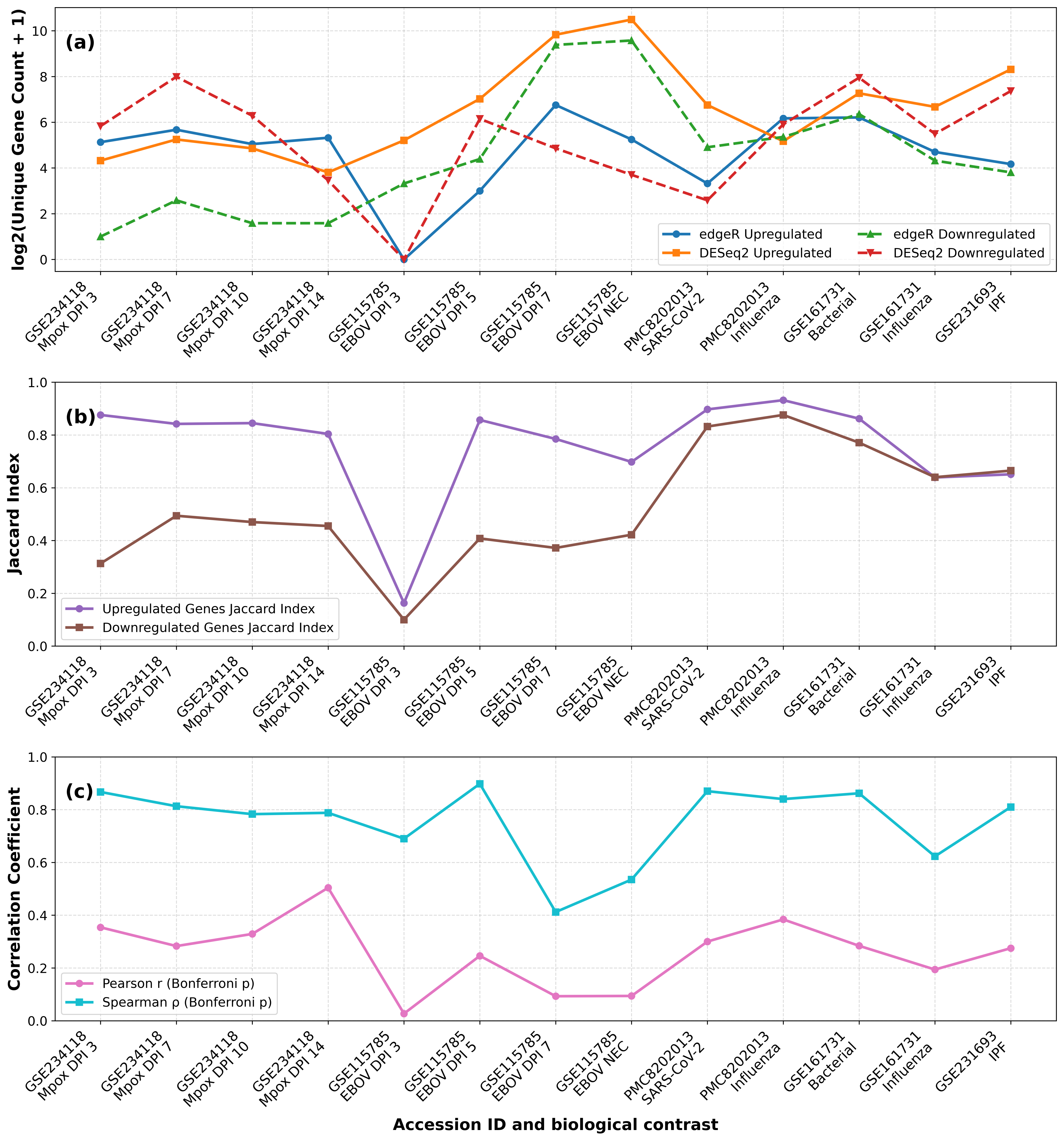}
\caption{\textbf{Comparison of \texttt{edgeR} and \texttt{DESeq2} across multiple biological contrasts.} 
Panel (a) shows the log$_2$-scaled number of uniquely identified upregulated and downregulated genes by each tool across 13 contrasts spanning viral, bacterial, and fibrotic conditions. 
Panel (b) displays the Jaccard index for upregulated and downregulated gene sets, indicating overlap between tools. 
Panel (c) shows Pearson and Spearman correlation coefficients computed for Bonferroni-adjusted $p$-values among common significant genes. 
Lines connecting points are included only as visual guides to aid comparison across contrasts and do not represent a continuous trend, interpolation, or fitted model. Dataset labels are shown as accession ID followed by biological contrast; infection timepoints are shown as DPI or NEC where applicable. The two influenza contrasts are labeled separately as PMC8202013 | Influenza and GSE161731 | Influenza.
\label{fig:Main-2}}
\end{figure}

Figure~\ref{fig:Main-2} summarizes the comparison of \texttt{edgeR} and \texttt{DESeq2} across 13 biological contrasts spanning viral, bacterial, and fibrotic conditions. Each panel quantifies different aspects of agreement and divergence in DEG calls between the two tools. Panel (a) displays the number of genes uniquely identified as significantly upregulated or downregulated by each tool, log$_2$-transformed for scale. Across most datasets, \texttt{DESeq2} identified a substantially larger number of unique upregulated genes, with pronounced differences observed in GSE115785 | EBOV DPI 7 and GSE115785 | EBOV NEC (e.g., 908 and 1,439 uniquely upregulated genes, respectively, compared to 107 and 37 for \texttt{edgeR}). The pattern was even more marked for downregulated genes: in GSE234118 | Mpox DPI 3 and GSE234118 | Mpox DPI 7, DESeq2 identified 56 and 253 uniquely downregulated genes respectively, while edgeR identified only 1 and 5. However, for EBOV contrasts (DPI 7 and NEC), edgeR identified far more uniquely downregulated genes (667 and 762) compared to DESeq2 (28 and 12), suggesting some context-specific reversal in sensitivity.

Panel (b) illustrates the Jaccard index between upregulated and downregulated gene sets from both tools across all datasets. The Jaccard index for upregulated genes was generally high (often exceeding 0.8), reflecting strong overlap between the two tools for the most transcriptionally active genes. However, downregulated gene sets showed much weaker agreement, with Jaccard indices ranging from 0.10 to 0.87. This discrepancy suggests that downregulated genes are less consistently detected across tools, likely due to lower signal strength or tool-specific modeling differences in shrinkage and dispersion estimation.

Panel (c) shows correlation coefficients (Pearson and Spearman) for Bonferroni-adjusted $p$-values among the common significant genes between tools. While Spearman correlations were consistently high (typically $>0.75$), indicating agreement in rank ordering, Pearson correlations were much lower, often below 0.4, and in one case (EBOV-DPI 3) dropped as low as 0.027. This indicates a non-linear relationship in adjusted significance levels despite overall agreement in gene ranking, and reinforces that tool-specific modeling may affect statistical inference even when fold-change estimates are aligned.

\begin{figure}[!ht]
\centering
\includegraphics[width=0.95\textwidth]{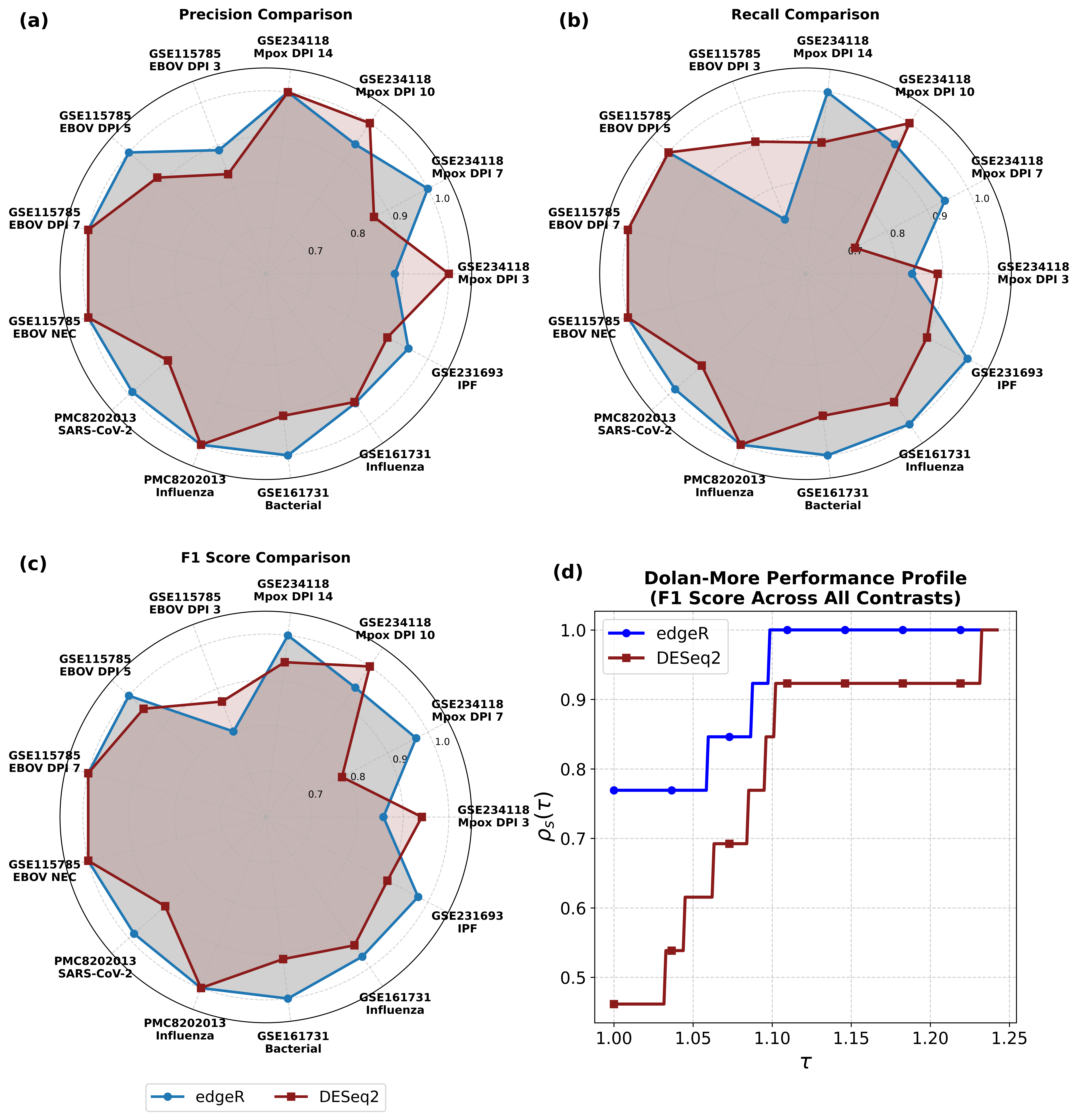}
\caption{\textbf{Classification performance of uniquely identified genes from \texttt{edgeR} and \texttt{DESeq2}.} 
Each gene set was evaluated using a logistic regression classifier trained on PC1 and PC2 from log-transformed expression values.
Panel (a) shows precision, (b) shows recall, and (c) shows F1 score for each dataset. 
Higher values indicate greater biological
separability of control versus treated samples. 
Panel (d) shows the Dolan-More profile of both tools based on F1 scores, summarizing
overall method robustness across all datasets.
\label{fig:Main-3}}
\end{figure}

Figure~\ref{fig:Main-3} evaluates the biological relevance of tool-specific significant genes identified by \texttt{edgeR} and \texttt{DESeq2} across 13 datasets via classification. Classification performance was assessed using precision, recall, and F1 score, shown respectively in panels (a), (b), and (c). Panel (a) shows that both tools achieved consistently high precision across datasets, often exceeding 0.9. \texttt{edgeR} outperformed \texttt{DESeq2} in 7 of the 13 contrasts, achieving perfect or near-perfect precision (1.000 or $>$0.98) in GSE234118 | Mpox DPI 7, GSE234118 | Mpox DPI 10, GSE234118 | Mpox DPI 14, and all GSE115785 | EBOV contrasts. In comparison, \texttt{DESeq2} had higher precision in GSE234118 | Mpox DPI 3, GSE231693 | IPF, and GSE161731 | Influenza. Both tools performed comparably well in PMC8202013 | SARS-CoV-2, GSE161731 | Bacterial, and the influenza contrasts.

Panel (b) presents recall values, showing that \texttt{edgeR} exhibited greater consistency across contrasts, with perfect recall (1.000) in 7 contrasts, particularly in GSE115785 | EBOV and inflammatory disease comparisons. \texttt{DESeq2} showed lower recall in GSE234118 | Mpox DPI 7 (0.722) and PMC8202013 | SARS-CoV-2 (0.903), though it matched \texttt{edgeR} in several other contrasts. Panel (c) summarizes classification performance via F1 score, capturing the trade-off between precision and recall. \texttt{edgeR} outperformed \texttt{DESeq2} in 9 out of 13 contrasts, achieving high F1 scores in GSE234118 | Mpox DPI 7 (0.971), GSE115785 | EBOV DPI 3 (0.800), and GSE231693 | IPF (0.976). \texttt{DESeq2}, while yielding generally strong results, showed more variability with notably lower F1 scores in GSE234118 | Mpox DPI 7 (0.788) and EBOV DPI 3 (0.870). Overall, \texttt{edgeR} provided more robust gene sets for classifying biological conditions.

Panel (d) shows the Dolan-More performance profile for both tools, providing a global benchmark of method consistency across all datasets. This method quantifies, for each tool, the fraction of datasets where its F1 score is within a factor $\tau$ of the best-performing method. \texttt{edgeR} achieved the highest F1 score in 10 of 13 datasets ($\rho(1) = 0.77$), with an average performance ratio of 1.02. In contrast, \texttt{DESeq2} was optimal in 6 datasets ($\rho(1) = 0.46$), with a slightly higher average ratio of 1.05. The Dolan-More curves illustrate that \texttt{edgeR} reaches $\rho(\tau) = 1$ faster, indicating greater consistency and reliability across a range of biological contrasts.

\subsection*{Pathway-level concordance of significant DEG sets}

We next evaluated whether differences in significant DEG sets between \texttt{edgeR} and \texttt{DESeq2} led to differences in downstream pathway enrichment results. For each dataset contrast, significant DEGs from each tool were analyzed using Hallmark and KEGG pathway collections. Enriched pathways were classified as \texttt{edgeR}-only, shared by both tools, or \texttt{DESeq2}-only, and pathway-level Jaccard similarity was used to quantify agreement between tools.

\begin{figure}[!ht]
\centering
\includegraphics[width=0.98\textwidth]{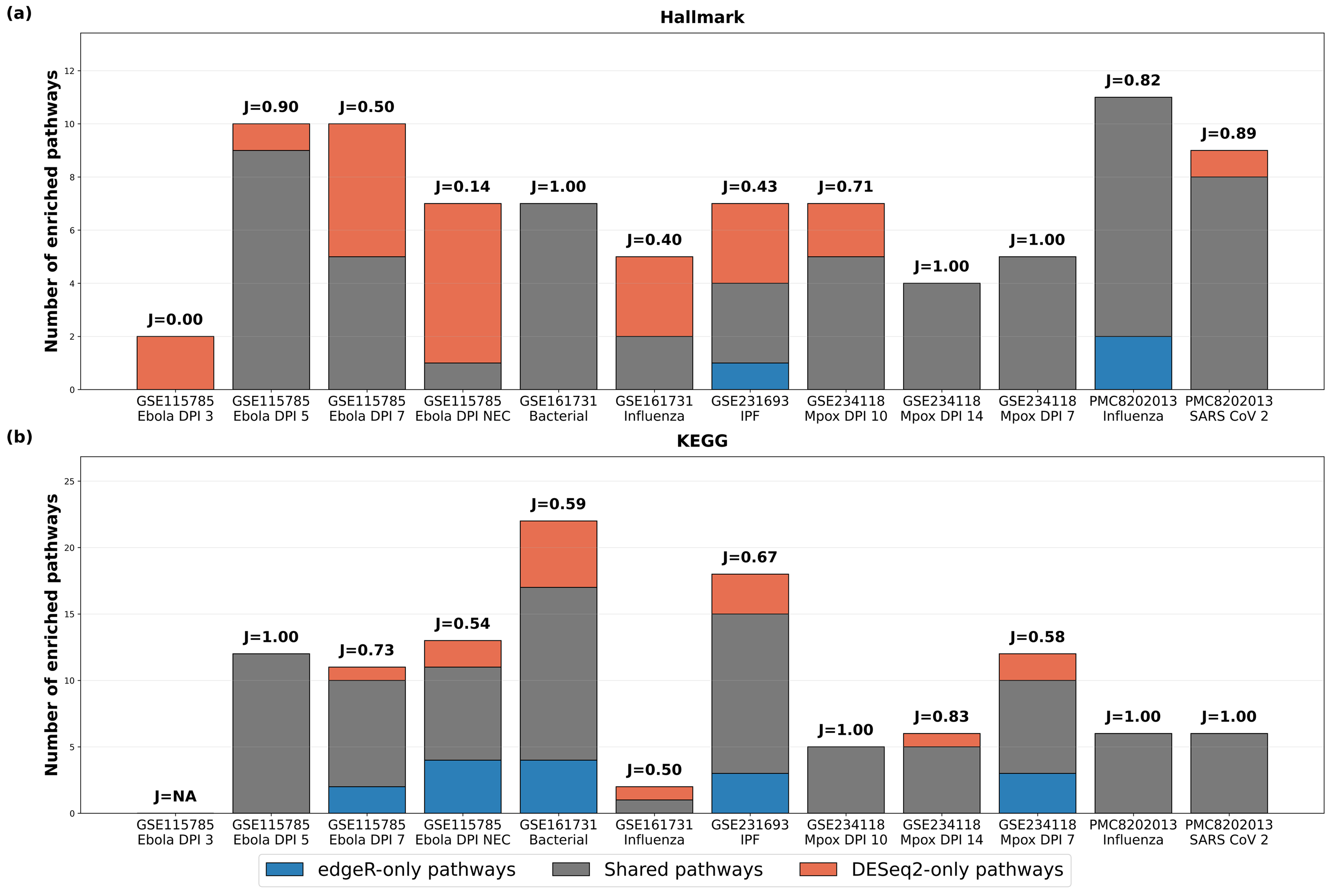}
\caption{\textbf{Pathway-level concordance between \texttt{edgeR} and \texttt{DESeq2} using significant DEG sets.}
Panel (a) shows Hallmark pathway enrichment results, and panel (b) shows KEGG pathway enrichment results. For each dataset contrast, enriched pathways were classified as \texttt{edgeR}-only, shared by both tools, or \texttt{DESeq2}-only using FDR \(< 0.05\). The value above each bar indicates the Jaccard similarity between the enriched pathway sets from the two tools. Higher Jaccard values indicate stronger pathway-level concordance.}
\label{fig:pathway_counts}
\end{figure}

Figure~\ref{fig:pathway_counts} summarizes enriched pathway counts across all evaluated contrasts. Hallmark enrichment results showed substantial pathway-level agreement in several contrasts, including cases where most enriched pathways were shared between \texttt{edgeR} and \texttt{DESeq2}. However, selected contrasts showed lower Jaccard similarity and a larger number of tool-specific enriched pathways. KEGG enrichment showed a similar pattern: several contrasts produced highly concordant pathway-level results, while others retained clear \texttt{edgeR}-only or \texttt{DESeq2}-only pathway enrichment. These results indicate that pathway-level conclusions can remain concordant even when DEG counts differ, but downstream pathway interpretation can still be tool-dependent in some contrasts.

To better visualize the contrasts with the largest pathway-level differences between tools, we selected representative comparisons with the strongest divergence in enriched pathway composition (Figure~\ref{fig:pathway_bubble}). This analysis highlights pathways that were uniquely enriched by \texttt{edgeR}, shared by both tools, or uniquely enriched by \texttt{DESeq2}. For both Hallmark and KEGG collections, the selected contrasts showed a mixture of shared and tool-specific pathway signals. Shared pathways often captured broad biological programs, whereas tool-specific pathways reflected differences in the significant genes prioritized by each method.

\begin{figure}[htbp]
\centering
\includegraphics[width=0.88\textwidth]{Figure2_selected_pathway_difference_bubbleplot.png}
\caption{\textbf{Selected pathway-level differences between \texttt{edgeR} and \texttt{DESeq2}.}
Panel (a) shows selected Hallmark contrasts, and panel (b) shows selected KEGG contrasts with the largest pathway-level differences between the tools. Bubble color indicates whether an enriched pathway was identified only by \texttt{edgeR}, shared by both tools, or identified only by \texttt{DESeq2}. Bubble size represents enrichment strength, quantified as \(-\log_{10}(\mathrm{FDR})\).}
\label{fig:pathway_bubble}
\end{figure}

\subsection*{Generalizability of tool-specific gene sets across independent studies}

We next evaluated whether tool-specific significant genes discovered in a subset of SARS-CoV-2 datasets could generalize to an unseen independent dataset. This analysis used four independent bulk RNA-seq datasets of SARS-CoV-2 infection and followed a leave-one-dataset-out validation design. In each fold, three datasets were used to identify reproducible tool-specific training signatures, and the remaining dataset was held out for validation. The procedure was repeated four times, so that GSE152418, GSE161731, GSE171110, and PMC8202013 each served once as the held-out test dataset.

\begin{figure}[htbp]
\centering
\includegraphics[width=0.98\textwidth]{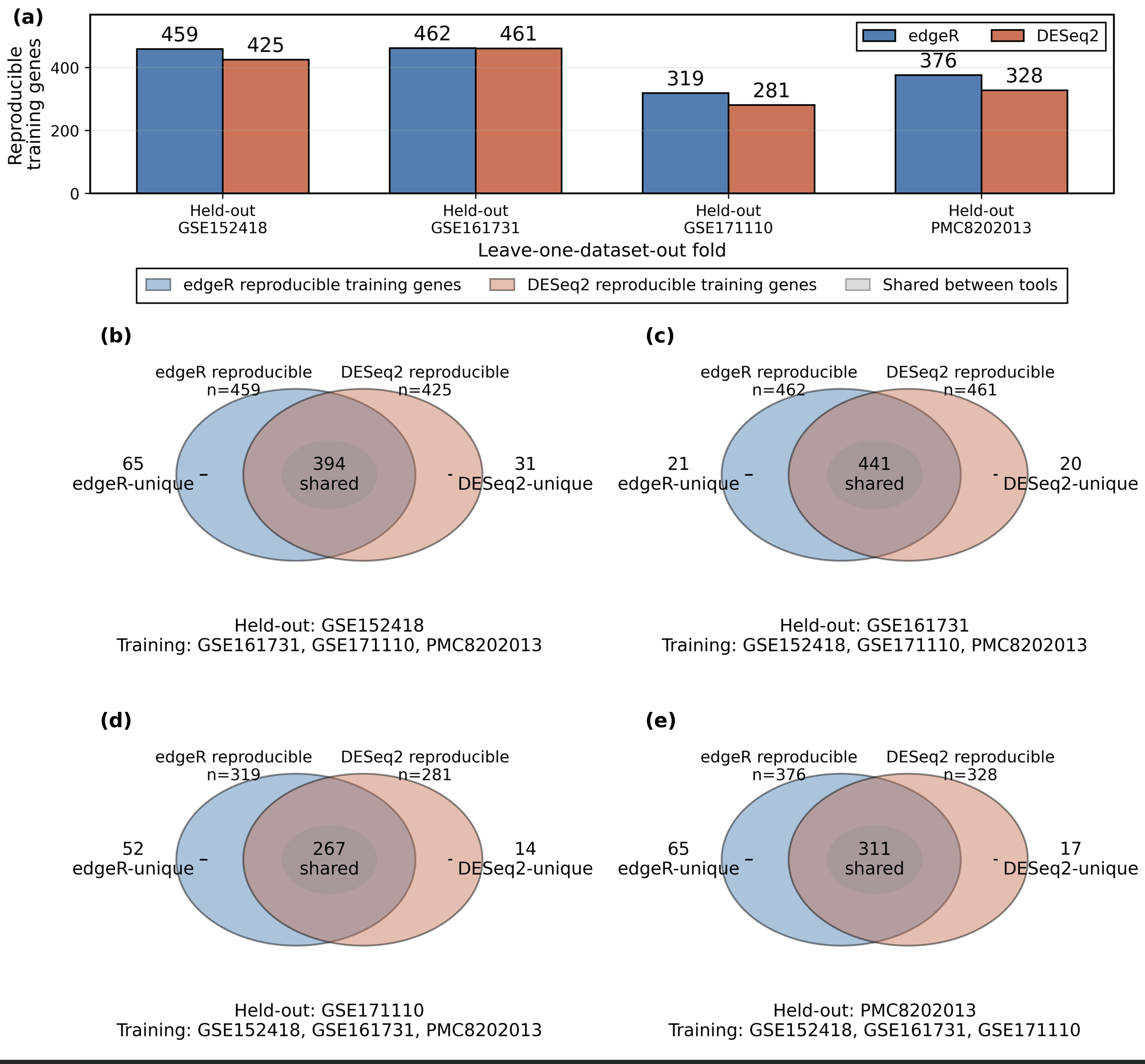}
\caption{\textbf{Training gene-selection procedure for leave-one-dataset-out cross-study validation.} 
Panel (a) shows the number of reproducible training genes identified by \texttt{edgeR} and \texttt{DESeq2} in each fold. A gene was considered reproducible for a given tool if it was significant in all three training datasets using FDR \(< 0.05\) and \(|\log_2 \mathrm{FC}| > 1\). 
Panels (b--e) show the decomposition of these reproducible training genes into \texttt{edgeR}-unique genes, genes shared by both tools, and \texttt{DESeq2}-unique genes for each held-out dataset. Unique gene counts are shown outside the Venn diagrams next to the corresponding tool, while shared gene counts are shown in the overlap. These tool-specific unique gene sets were subsequently evaluated in the corresponding held-out dataset.}
\label{fig:training_gene_selection}
\end{figure}

Figure~\ref{fig:training_gene_selection} summarizes the training gene-selection procedure used in each fold. In panel (a), reproducible training genes were defined as genes that were significant in all three training datasets for a given tool using FDR \(< 0.05\) and \(|\log_2 \mathrm{FC}| > 1\). When GSE152418 was held out, \texttt{edgeR} and \texttt{DESeq2} identified 459 and 425 reproducible training genes, respectively. When GSE161731 was held out, the corresponding counts were 462 and 461; when GSE171110 was held out, they were 319 and 281; and when PMC8202013 was held out, they were 376 and 328.

Panels (b--e) show how these reproducible training genes were divided into genes unique to \texttt{edgeR}, genes shared by both tools, and genes unique to \texttt{DESeq2}. Across all four folds, most reproducible training genes were shared between the two tools. However, each fold also retained tool-specific signatures. For the GSE152418, GSE161731, GSE171110, and PMC8202013 held-out folds, the numbers of \texttt{edgeR}-unique genes were 65, 21, 52, and 65, respectively, while the corresponding numbers of \texttt{DESeq2}-unique genes were 31, 20, 14, and 17. The numbers of shared reproducible genes were 394, 441, 267, and 311 across these folds. 

\begin{figure}[htbp]
\centering
\includegraphics[width=0.95\textwidth]{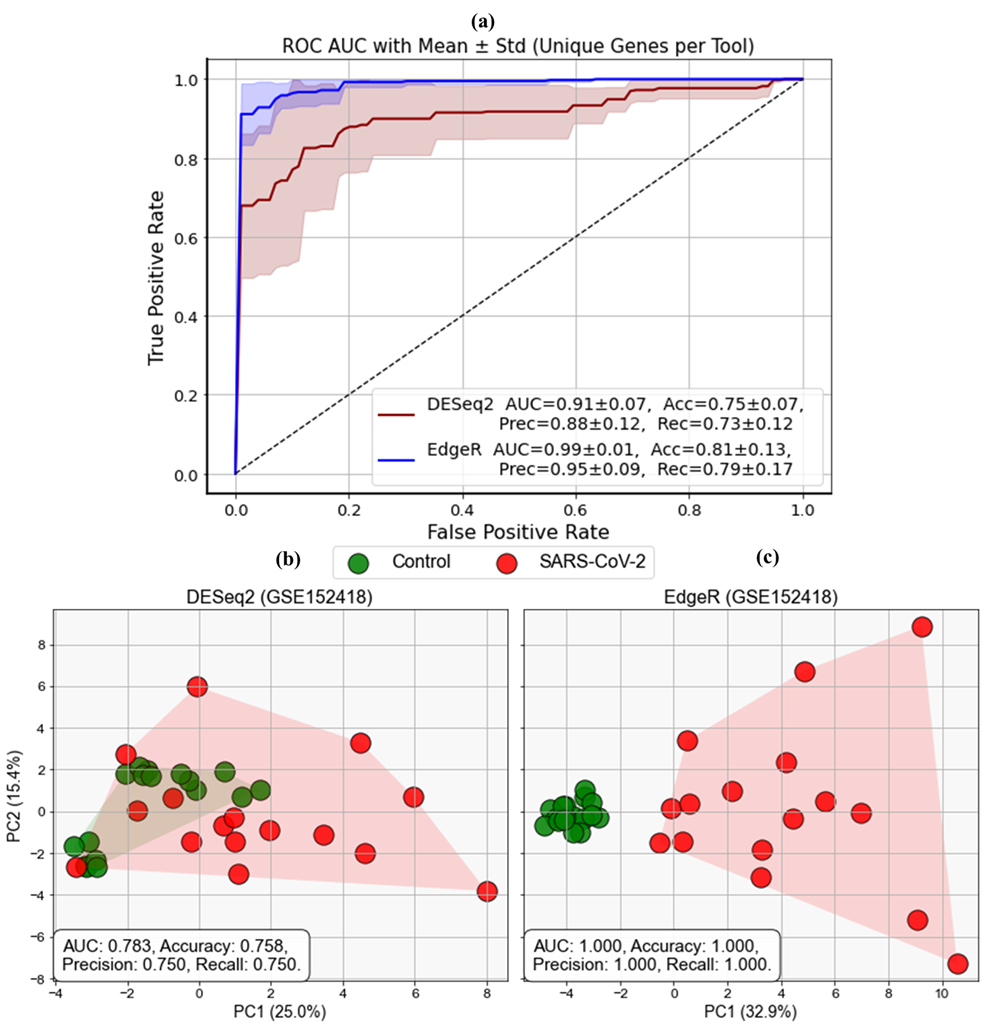}
\caption{\textbf{Cross-study generalizability of uniquely significant genes from \texttt{edgeR} and \texttt{DESeq2}.} 
(a) Mean ROC curves across four independent SARS-CoV-2 datasets, with shaded regions representing $\pm 1$ standard deviation of the true positive rate (TPR) at each false positive rate (FPR). Metrics in the legend summarize mean AUC, accuracy, precision, and recall with standard deviation. 
(b) PCA-based classification using \texttt{DESeq2}-specific genes from training datasets applied to test set GSE152418, yielding moderate separation (AUC = 0.783). 
(c) Corresponding classification using \texttt{edgeR}-specific genes for the same test set, yielding perfect separation (AUC = 1.000).
\label{fig:Main-4}}
\end{figure}

After defining the reproducible tool-specific signatures shown in Figure~\ref{fig:training_gene_selection}, Figure~\ref{fig:Main-4} evaluates whether these signatures remained predictive in the corresponding held-out datasets. This analysis assessed whether genes uniquely and reproducibly identified by each tool captured biological signals that generalized beyond the datasets used for gene selection.

Panel (a) presents the average ROC curves with shaded bands indicating $\pm 1$ standard deviation of the true positive rate (TPR) across folds. \texttt{edgeR} achieved a mean AUC of $0.99 \pm 0.01$, accuracy of $0.81 \pm 0.13$, precision of $0.95 \pm 0.09$, and recall of $0.79 \pm 0.17$. In contrast, \texttt{DESeq2} yielded lower performance with a mean AUC of $0.91 \pm 0.07$, accuracy of $0.75 \pm 0.07$, precision of $0.88 \pm 0.12$, and recall of $0.73 \pm 0.12$.

Panels (b) and (c) show representative classification performance on the test dataset GSE152418. For \texttt{DESeq2}-specific genes (panel b), the model yielded AUC = 0.783, with accuracy, precision, and recall all equal to 0.75. In contrast, using \texttt{edgeR}-specific genes (panel c) resulted in perfect separation, with AUC, accuracy, precision, and recall all equal to 1.000.

\section*{Discussion}

\subsection*{Sensitivity to sample size and robustness to outliers}

The semi-simulated experiments presented in Figure~\ref{fig:Main-1} highlight distinct performance profiles of \texttt{edgeR} and \texttt{DESeq2} with respect to sample size and robustness to outliers. \texttt{DESeq2} consistently identified more DEGs than \texttt{edgeR} across all tested sample sizes, with particularly notable differences at low $n$. This behavior may reflect \texttt{DESeq2}’s use of empirical Bayes shrinkage and regularization, which improves variance estimation under limited replication. While such conservatism helps stabilize inference, it also introduces a more inclusive DEG set in small-$n$ settings, potentially increasing the risk of false positives.

As sample size increased, the DEG sets produced by both tools converged in both size and content. Directional overlap analysis demonstrated that \texttt{edgeR} recovered a growing proportion of \texttt{DESeq2}-identified genes as statistical power increased. By $n=45$, both tools achieved over 95\% mutual overlap, suggesting that differences in underlying models and statistical testing become less consequential when replication is sufficient. These findings suggest that, for well-powered studies, both tools offer highly comparable DEG discovery performance.

Robustness analysis under sample swapping (Figure 1j) revealed that both tools degrade predictably in the presence of controlled contamination, with Jaccard similarity decreasing as the number of outlier samples increased. However, \texttt{DESeq2} maintained slightly higher mean similarity scores and lower variability across replicates at all levels of perturbation. This suggests that \texttt{DESeq2} offers marginally greater robustness to moderate outlier effects.

Together, these findings indicate that \texttt{DESeq2} is more sensitive to detecting significant genes at small sample sizes, though whether this increased sensitivity reflects true biological signal or introduces additional noise is a question discussed in the next two sections. As sample size increases, both tools become more consistent, with convergence in both DEG count and content.

\subsection*{Classification performance of uniquely identified gene sets within the discovery dataset}

The classification-based analysis presented in Figures~\ref{fig:Main-2} and ~\ref{fig:Main-3} offers a performance-driven assessment of the biological validity of genes uniquely identified by \texttt{edgeR} and \texttt{DESeq2}. By training logistic regression models using PC1 and PC2 derived from the expression of uniquely significant genes, we evaluated each tool’s ability to recover gene sets that effectively discriminate between control and treated samples. This framework serves as a proxy for evaluating the ``trueness’’ of significant genes in the absence of an external ground truth.

While both tools demonstrated high precision across most datasets (Figure~\ref{fig:Main-3}a), \texttt{edgeR} consistently achieved equal or superior precision in more than half of the contrasts. This indicates that, despite identifying fewer unique DEGs than \texttt{DESeq2}, the genes it does report tend to be more predictive and less noisy. The recall results in Figure~\ref{fig:Main-3}b further reinforce this observation, with \texttt{edgeR} exhibiting stronger sensitivity in multiple datasets, especially in EBOV and inflammatory disease contrasts.

The F1 score results shown in Figure~\ref{fig:Main-3}c provide an integrated view of classification performance, balancing both precision and recall. \texttt{edgeR} outperformed \texttt{DESeq2} in 9 out of 13 contrasts, achieving notably high scores in GSE234118 | Mpox DPI 7 (0.971) and EBOV DPI 3 (0.800), and GSE231693 | IPF (0.976). In contrast, \texttt{DESeq2} showed larger fluctuations in F1 performance, with reduced scores particularly in datasets where it identified many unique genes that ultimately contributed less to sample separability.

These trends suggest that some of the uniquely identified genes from \texttt{DESeq2}, especially in low-sample or highly variable settings, may include false positives. This finding aligns with earlier observations (Figure~\ref{fig:Main-1}) that \texttt{DESeq2} is more inclusive in small-$n$ contexts, potentially increasing sensitivity at the expense of specificity.

The Dolan-More profile in Figure~\ref{fig:Main-3}d further highlights the comparative consistency of each method across diverse datasets. \texttt{edgeR} attained the highest F1 score in a majority of contrasts and exhibited more stable performance overall, reinforcing its reliability in identifying gene sets with meaningful classification potential.

In summary, while \texttt{DESeq2} tends to identify more significant genes, \texttt{edgeR} provides gene sets that are, on average, more predictive of biological condition when evaluated via classification. This suggests that increased sensitivity in DEG calling, especially in smaller datasets, may not always correspond to biological relevance, and highlights the importance of downstream validation when interpreting differential expression results.

\subsection*{Pathway enrichment concordance between tools}

The pathway enrichment concordance analysis provides an additional view of how tool choice affects downstream interpretation (Figure~\ref{fig:pathway_bubble} and Figure~\ref{fig:pathway_counts}). Although the DEG-level analyses showed that \texttt{edgeR} and \texttt{DESeq2} can differ in the number and identity of significant genes, the pathway-level results showed that many biological interpretations remained concordant between tools. This suggests that, in several contrasts, the dominant pathway-level signals are robust to differences in DEG calling.

However, pathway-level agreement was not uniform across all contrasts. Some comparisons showed clear tool-specific enriched pathways, indicating that differences in significant DEG sets can affect downstream pathway interpretation. This is important because pathway analysis is frequently used to summarize DEG results into biological themes. Therefore, even when two tools identify broadly similar biological signals, tool-specific DEGs can still alter which pathways are emphasized.

These findings refine the interpretation of the DEG-level results. Tool choice can influence downstream enrichment results, but the extent of this influence depends on the dataset, the biological contrast, and the pathway collection used. Thus, pathway-level concordance should be evaluated alongside DEG counts, classification performance, and cross-study validation when comparing differential expression tools.

\subsection*{Generalizability of tool-specific gene sets across independent studies}

The cross-study validation results presented in Figure~\ref{fig:training_gene_selection} and Figure~\ref{fig:Main-4} offer compelling evidence that the biological signal captured by tool-specific gene sets differs in generalizability and predictive power. Although both \texttt{edgeR} and \texttt{DESeq2} are widely accepted for differential gene expression analysis, the present results highlight clear differences in the robustness of the gene sets they identify when transferred to independent datasets.

One of the most important findings is that \texttt{edgeR}-specific genes yielded significantly higher classification performance than those uniquely identified by \texttt{DESeq2}. This was consistent across multiple folds of leave-one-out cross-study validation and was further exemplified by the perfect classification obtained in the representative GSE152418 test case. These results indicate that \texttt{edgeR} not only identifies fewer unique genes but that these genes are more likely to represent consistent, transferable biological signals rather than study-specific artifacts.

In contrast, the performance drop observed with \texttt{DESeq2}-specific genes suggests that some of its identified features may be more reflective of noise or dataset-specific variance. This aligns with previous observations from our semi-simulation and classification experiments, where \texttt{DESeq2} exhibited increased sensitivity but also greater variability in performance. It is plausible that \texttt{DESeq2}'s regularized dispersion estimation and more inclusive thresholding increase the likelihood of capturing subtle but less reproducible patterns, particularly in smaller or noisier datasets.

The contrast between the two tools in this context reveals a fundamental trade-off. \texttt{DESeq2} appears more permissive and sensitive, which may be advantageous for exploratory analyses or hypothesis generation but potentially at the cost of precision and reproducibility. On the other hand, \texttt{edgeR} employs stricter criteria that may reduce the total number of reported DEGs but improve their specificity and cross-dataset stability. In the context of biomarker discovery or translational applications where reproducibility is paramount, the conservative profile of \texttt{edgeR} may be preferable.

Furthermore, these findings reinforce the importance of integrating downstream classification or validation frameworks into DEG analysis pipelines. The number of DEGs alone is not a sufficient metric for evaluating tool performance; rather, the ability of these genes to generalize across biological contexts and datasets is a more meaningful benchmark. The consistent superiority of \texttt{edgeR} in our cross-study framework underscores its capacity to identify gene sets that are not only statistically significant but biologically informative and generalizable.

In sum, while both tools are capable of capturing relevant gene expression changes, \texttt{edgeR} provides more reliable and transferable gene sets, particularly in applications demanding high reproducibility. These results advocate for tool selection to be guided not only by statistical properties but also by the intended downstream use of the identified genes.

\subsection*{Integration with prior benchmarks and implications for tool choice}

Taken together, our results extend previous benchmarking studies by showing how statistical differences between \texttt{edgeR} and \texttt{DESeq2} translate into downstream biological utility. Prior work has already shown that RNA-seq DGE tools can differ in DEG yield, type I error behavior, robustness to perturbation, and sensitivity to sample size \citep{seyednasrollah2015comparison, zhang2014comparative, stupnikov2021robustness, li2022exaggerated, li2022evaluation, cui2021super}. The main contribution of the present study is that we evaluated what happens after tool-specific genes are identified: whether those genes separate biological groups, whether they lead to concordant pathway-level interpretations, and whether they generalize across independent datasets. This downstream evaluation is important because a DEG list can appear statistically significant yet still be less useful for classification, pathway interpretation, or transfer to an external cohort.

The sample-size analysis provides one point of agreement with previous work and one important extension. Consistent with earlier studies, we found that tool agreement was lowest at smaller sample sizes and increased as replication improved. However, the direction of the difference in our data was informative: \texttt{DESeq2} generally identified more DEGs than \texttt{edgeR} at smaller sample sizes, whereas the two tools became highly concordant in the full RSVB dataset. This suggests that, when sample size is adequate and the biological signal is strong, the practical difference between the two tools may become smaller. In contrast, when sample size is limited, tool-specific modeling choices can have a larger effect on the resulting DEG list. This finding is consistent with the broader literature, but our analysis further shows that early differences in DEG yield can carry into downstream classification and external validation.

The classification results provide a functional perspective on the false-positive concern raised in previous studies. We did not directly estimate gene-level FDR using permutation or known ground truth, so our results should not be interpreted as proving that a specific set of \texttt{DESeq2}-specific genes are false positives. Instead, our analysis asks a complementary question: when each tool identifies genes that the other tool does not, which set better captures biological separation? Across 13 contrasts, \texttt{edgeR}-specific genes more often produced higher F1 scores and more consistent classification performance. This pattern suggests that the larger \texttt{DESeq2}-specific gene sets observed in some contrasts may include genes with weaker biological separability, whereas the more restricted \texttt{edgeR}-specific sets more often captured signal that distinguished the relevant biological groups. This result is consistent with prior reports that larger DEG lists may include less stable or less reliable signals, but it adds an explicit classification-based evaluation that most earlier studies did not include.

The pathway enrichment analysis adds a second layer of interpretation. Previous studies have shown that tool-specific DEG lists can produce different functional conclusions, and Li et al. showed that spurious DEG calls can even generate biologically plausible pathway enrichment \citep{li2022exaggerated}. In the present study, many contrasts showed strong pathway-level agreement between \texttt{edgeR} and \texttt{DESeq2}, indicating that major biological programs can remain stable even when DEG lists are not identical. At the same time, selected contrasts showed tool-specific enriched pathways, which means that downstream interpretation can still depend on the DGE tool. This distinction is important: tool choice does not always change the biological conclusion, but it can change which pathways are emphasized, especially when tool-specific genes drive enrichment results.

The cross-study SARS-CoV-2 validation provides the strongest practical distinction between the two tools in this study. Genes uniquely and reproducibly identified by \texttt{edgeR} across training datasets produced higher AUC, precision, and recall in held-out datasets than genes uniquely and reproducibly identified by \texttt{DESeq2}. This result is not simply another measure of DEG overlap or statistical significance; it directly evaluates transferability across independent studies. In this sense, our findings extend previous benchmarking work by showing that tool-specific gene sets can differ in external predictive value. For applications such as biomarker discovery, signature development, or translational studies, this type of external performance may be more informative than DEG count alone.

Overall, the combined evidence suggests a practical interpretation of \texttt{edgeR} and \texttt{DESeq2}. \texttt{DESeq2} may be useful for broad exploratory discovery because it often identifies a larger candidate set, especially in smaller-sample settings. However, larger DEG yield should be followed by additional checks, including pathway concordance, sensitivity to outliers, and validation in independent data. \texttt{edgeR}, in our analyses, produced tool-specific gene sets that were more often predictive and transferable, which makes it attractive when reproducibility and downstream validation are priorities. This does not mean that \texttt{edgeR} is universally superior or that \texttt{DESeq2} is unreliable. Prior studies show that both tools can be affected by dataset structure, outliers, filtering, and model assumptions. Rather, the practical message is that \texttt{DESeq2} may be better suited for sensitivity-oriented candidate generation, while \texttt{edgeR} may be preferable when the primary goal is a smaller, more reproducible gene set with stronger downstream performance.

A useful strategy is therefore to treat shared genes and shared pathways from both tools as a robust core signal, while interpreting tool-specific genes according to the goal of the study. If the goal is hypothesis generation, tool-specific genes from either method may be worth exploring. If the goal is reproducible classification or biomarker development, tool-specific genes should be evaluated using downstream separability and external validation, as done here. This study therefore supports a broader view of DGE benchmarking: the best tool is not necessarily the one that identifies the most genes, but the one whose genes remain stable, interpretable, and useful beyond the discovery dataset.

\section*{Conclusions}

This study presents a systematic and multifaceted comparison of two widely used RNA-Seq differential expression tools, \texttt{edgeR} and \texttt{DESeq2}, across a broad spectrum of analytical challenges, including sensitivity to sample size, robustness to outliers, classification-based evaluation of unique gene sets, and cross-study generalizability. Our findings offer nuanced insights into the trade-offs and strengths of each method, providing practical guidance for tool selection in transcriptomic studies.

We find that \texttt{DESeq2} often identifies more differentially expressed genes (DEGs), especially in small sample settings. This increased sensitivity, however, comes with greater susceptibility to noise and less consistent performance when applied to independent datasets. In contrast, \texttt{edgeR} identifies fewer DEGs but exhibits more conservative and stable behavior, particularly as sample size increases. The two tools converge in performance under well-powered designs, with over 95\% mutual overlap in DEG sets by $n=45$, indicating that replication reduces methodological divergence.

Classification-based evaluation of uniquely identified genes reveals that \texttt{edgeR}-specific gene sets tend to be more predictive of biological condition, achieving higher F1 scores in the majority of cases. This suggests that \texttt{edgeR} is more effective at prioritizing biologically informative genes with strong signal-to-noise ratios, while \texttt{DESeq2}'s greater inclusiveness may introduce more marginal or context-specific features.

Pathway enrichment analysis further showed that DEG-level differences do not always translate into divergent downstream biological interpretation. Hallmark and KEGG enrichment results were concordant across many contrasts, although selected comparisons retained tool-specific enriched pathways, indicating that pathway-level validation can provide an important complement to DEG-level benchmarking.

Cross-study validation further reinforces these distinctions. Gene sets uniquely discovered by \texttt{edgeR} generalized more effectively across independent SARS-CoV-2 datasets, achieving nearly perfect classification performance in multiple test cases. In contrast, \texttt{DESeq2}-specific gene sets demonstrated lower reproducibility and weaker discriminatory power when transferred to unseen datasets. These findings highlight \texttt{edgeR}'s superior utility for biomarker discovery, where robustness and reproducibility across cohorts are essential.

Collectively, our results emphasize that no single tool is universally superior; rather, each has context-dependent advantages. For studies focused on hypothesis generation or underpowered designs, \texttt{DESeq2}'s sensitivity may be desirable. However, when prioritizing specificity, cross-dataset reproducibility, or translational applications, \texttt{edgeR}'s conservative and robust profile makes it a more reliable choice. The addition of pathway-level concordance analysis highlights that tool selection should be evaluated across both gene-level and pathway-level outcomes. These insights advocate for tailoring tool selection to the study's design constraints and downstream objectives, and underscore the value of incorporating biological validation frameworks into differential expression analyses.

\section*{Limitations and Future Work}

This study has several limitations. First, our comparison was limited to two tools, \texttt{edgeR} and \texttt{DESeq2}, using their default pipelines. Future work should assess additional methods and explore the impact of parameter tuning. Second, while we evaluated biological relevance via classification and cross-study validation, external ground truth data (e.g., qPCR validation or functional assays) were not available, which limits definitive biological interpretation. Expanding to other disease models and incorporating validation strategies will help generalize and strengthen these findings.

\section*{Data and Code Availability}

All datasets used in this study are publicly available from the NCBI Gene Expression Omnibus (GEO) or PubMed Central (PMC): RSVB (GSE196134) \citep{anderson2024differential}, Mpox (GSE234118) \citep{aid2023mpox}, EBOV (GSE115785) \citep{cross2018comparative}, bacterial pneumonia and influenza (GSE161731) \citep{mcclain2021dysregulated}, idiopathic pulmonary fibrosis (GSE231693) \citep{jia2023interleukin}, and SARS-CoV-2 datasets including GSE152418 \citep{arunachalam2020systems}, GSE161731 \citep{mcclain2020dysregulated}, GSE171110 \citep{levy2021cd177}, and PMC8202013 \citep{bibert2021transcriptomic}. Detailed sample design and application of each dataset in this study are summarized in Table~\ref{tab:dataset_summary}. Codes are available at: \url{https://github.com/MostafaRezapour/Tool-Choice-Matters}


\section*{Author Contributions}
M.R. conceived the study, performed the analyses, and wrote the manuscript.

\section*{Competing Interests}
The authors declare no competing interests.

\bibliography{reference}

\end{document}